\documentclass[preprint,12pt]{elsarticle}
\usepackage{epsfig}
\usepackage{subfig}
\usepackage{amssymb}
\usepackage{multirow}
\journal{Astroparticle Physics}
\begin{document}
\begin{frontmatter}
\title{Simulation of proton-induced and iron-induced extensive air showers at extreme energies}
\author{Do Thi Hoai\corref{cor1}}
\ead{dohoai\_1987@yahoo.com}
\author{Pham Ngoc Diep, Pierre Darriulat,\\
Pham Tuan Anh, Pham Ngoc Dong, Nguyen Van Hiep, Pham Thi Tuyet Nhung and Nguyen Thi Thao}
\address{VATLY/INST, 179 Hoang Quoc Viet, Cau Giay, Ha Noi, Vietnam}
\cortext[cor1]{Corresponding author}
\begin{abstract}
The development of extensive air showers at extreme energies is studied using a simulation model much simpler and cruder, but also more transparent and flexible, than existing sophisticated codes. Evidence for its satisfactory performance is presented. As an illustration, shower elongation rates are evaluated in the $10^{18}$ to $10^{20}$ eV region and compared with recently published data. Lateral distribution functions of both muons and electrons/photons are also briefly discussed. Reliable results are obtained in the comparison between proton-induced and iron-induced showers.
\end{abstract}
\end{frontmatter}

% \linenumbers
\section{Introduction}
Cosmic rays are atomic nuclei that give, together with photons and magnetic fields, an important contribution to the energy balance of the interstellar medium. Their flux covers 32 orders of magnitude over 12 orders of magnitude in energy, with a power law of index $32/12 \sim 2.7$. They are dominated by protons; the relative abundance of nuclei of different species is similar to that found in the interstellar medium. The low energy end of the spectrum is associated with solar emissions that are prevented to reach the Earth by the shielding effect of the geomagnetic field. On Earth, most cosmic rays are of galactic origin and are accelerated in the shells of young Super Nova Remnants by a mechanism of Diffusive Shock Acceleration. The high energy end of the spectrum, one speaks of Ultra High Energy Cosmic Rays (UHECR), is of extragalactic origin and is currently the subject of extensive studies \cite{1}. It is cut-off, in the region of $\sim$10$^{20}$ eV, by the onset of photoproduction on the Cosmic Microwave Background \cite{2}, implying that possible sources should not be farther away from the Earth than some 50 to 100 Mpc. If the mechanism of acceleration is the same as for galactic cosmic rays, the sources must be host to very large shock fronts, such as the environment of Active Galactic Nuclei (AGN) may provide. However, a reliable identification of the sources remains today an open question. There is evidence from the Pierre Auger Observatory (PAO) for a positive correlation with galaxies and AGNs in the nearby Universe \cite{3}, in particular with Cen A, the closest AGN to Earth; but the association of UHECR showers with such celestial objects is not as sharp as one could expect if protons were dominant and if intergalactic magnetic fields were small enough. A possible explanation might be that at such high energies, protons are no longer dominant but leave the place to more massive nuclei such as iron: their large electric charges would result in important magnetic deflections and blur the image of the sources. Indeed, there are indications from the PAO that such is the case \cite{4} but the question is not yet settled.

Settling such questions is difficult because the UHECR rate is very small $-$in the equivalent of four years of operation, the PAO has collected no more than about hundred UHECRs of high enough energy for having a chance to identify their source$-$ and because the detection of UHECRs is an indirect process: what is detected is not the primary cosmic ray but the shower which it induces by interacting with the Earth atmosphere. Understanding how such showers develop, in particular how do showers initiated by different nuclei differ, is therefore a central problem of UHECR physics.  

Many authors have conceived and written simulation codes that aim at giving as precise and reliable as possible a description of the development of extensive air showers \cite{5}. However, such a task faces numerous difficulties, both conceptual and technical. 

The lack of relevant data is a major drawback. The closest available information is from the Large Hadron Collider (LHC) \cite{6}, at CERN, where proton-proton collisions are studied at an equivalent cosmic ray energy of 2.5 10$^{16}$ eV, 4000 times lower than the UHECR reach of 10$^{20}$ eV, and ion-ion collisions at 1.38 TeV per nucleon. Most of this information is on the central rapidity region because the LHC and its detectors are at rest in the centre of mass system of the collision; but in the case of cosmic ray physics, the detector is at rest in the rest frame of the target and it is the forward rapidity region, of difficult access to accelerators, which is relevant. Most of the UHECR shower development is governed by pion-air interactions, for which the only available accelerator data are at very low energies. How to extrapolate available knowledge from protons to air, from protons to pions, from central to forward rapidities, from LHC energies to centre-of-mass energies nearly two orders of magnitude larger is the task of the modelist who can only rely on his or her judgement and on the guidance provided by our current understanding of the underlying particle physics.

Technical difficulties result from the gigantic multiplicity of secondary particles in a UHECR shower: several billions, making it impossible to follow the details of their history within manageable computing times. Yet, in spite of this complexity, the global behaviour of a UHECR shower can be (and is) described in terms of a remarkably small number of parameters: four parameters to describe the longitudinal profile and a few more to describe the lateral profile on ground (one speaks of a lateral distribution function). All these parameters are found to vary slowly and nearly linearly with the logarithm of energy, and important scaling features can be unravelled.

In particular, such is the case of the depth at which the shower reaches maximum, traditionally called $X_{max}$, an important tracer of the nature of the primary cosmic ray. Indeed, the main difference between a shower induced by a primary proton and, say, by a primary iron nucleus is that the latter starts developing much earlier than the former and the associated $X_{max}$ is significantly smaller and fluctuates less around its mean value. The reason is simple: to the extent that an iron nucleus may be seen as a collection of 56 independent nucleons, and as the iron-air interaction cross-section is typically four times larger than the proton-air interaction cross-section, the first interaction will occur much higher in the upper atmosphere, and produce a much larger multiplicity of secondaries, in the case of a primary iron nucleus than in the case of a primary proton. Another very important consequence of this remark is that the fate of a shower is decided in the first interaction of the primary cosmic ray with the upper atmosphere and in those of the first generation of secondaries. What happens after implies such a large number of secondaries that it can be described statistically. In particular the mean value of $X_{max}$ and its root-mean-square deviation with respect to the mean, $Rms(X_{max})$, are essentially defined by the first interaction and those of the second generation.

Having in mind the above considerations, the present article describes a very simple model of UHECR shower development that aims at being more rapid, more transparent and more flexible than available sophisticated codes \cite{5}. This is at the price of extreme crudeness: the present model does not have the ambition to compete with these other codes, but simply to provide some complementary information. 

\section{Shower development}
\subsection{General strategy}
The method used here \cite{7} consists in following the development of the sub-shower induced by a secondary only when its energy exceeds some predefined fraction $f$ of the primary energy. When it does not, one uses instead a parameterised description of the sub-shower, which makes it unnecessary to follow the details of its subsequent development. The main argument in favour of such an approach, which treats precisely and reliably the first interactions taking place in the development of the shower, is that the fluctuations observed in the development of showers induced by primaries of a same nature and of a same energy are dominated by the very first interactions.

In a low energy range, 1 to 10$^3$ GeV, proton-induced showers are simulated without making use of any parameterisation of the hadronic sub-showers but making full use of the parameterisation of photon showers introduced in Section 2.2. In this energy range, the number of shower particles is small enough to follow each charged pion separately while keeping the computing time reasonable. At each node of a grid in energy, altitude above ground and zenith angle, the longitudinal profile and the lateral distribution functions of electron/photons and of muons are parameterised. The parameters are evaluated for each shower and their mean values are calculated. Once this is completed, shower parameterisations can be performed by interpolation of the parameters between the nodes. 
 
In a second phase, one calculates the parameters in the high energy range, above 10$^3$ GeV. One proceeds by iterations, in steps of half a unit of $lgE$, to extend the grid to higher energies. In this second step, one only follows charged secondaries having energies in excess of a fraction $f$ of the primary energy, and replaces each lower energy interacting pion by a parameterised sub-shower.

Many simplifications are being made in the description of the hadronic interactions, the most important being the assumption that all produced secondaries are pions. This is far from being the case; there are in particular an important number of kaons among the secondaries. Moreover, many pions are decay products of resonances. It is nevertheless reasonable to expect that the all-pion approximation can be used to describe reality, possibly at the price of adjusting parameters such as the pion decay time in an {\it ad hoc} manner.
 
The longitudinal development of the showers requires a description of the atmospheric pressure and of the electromagnetic interactions of charged particles with the atmosphere, causing energy losses and multiple Coulomb scattering. An exponential dependence of the atmospheric pressure $p$ as a function of altitude $z$ of the form $p=p_0exp(-z/\Delta z)$ has been retained, using $\Delta z=6.83$ km and $p_0=1100$ g/cm$^2$, which gives a good description of standard atmospheric profiles \cite{7}.
   
Two kinds of energy losses are taken into account: ionization losses and radiation losses. They are supposed to be the same when the incident energy $E$ is equal to the critical energy $E_{crit}$ taken as input parameter. Their precise forms are given in Reference 7. Multiple scattering in a slice of $x$ g/cm$^2$ is calculated \cite{7} using a mean transverse momentum kick of $13.6\sqrt{2x/X_0}$ MeV where $X_0$ is the radiation length in air, 36.7 g/cm$^2$. 

\subsection{Electromagnetic showers}
A large number of neutral pions are produced in the development of extensive air showers, of the order of one third of all secondaries. Neutral pions decay almost instantly in a pair of photons, which initiate electromagnetic showers and do not any longer contribute to the development of the hadronic shower. The method sketched in the preceding sub-section has been applied successfully to the longitudinal development of electromagnetic showers, in particular to the study of ultra high energy phenomena such as the LPM (Landau-Pomeranchuk-Migdal) and Perkins effects \cite{8}.
 
Two features make such treatment particularly simple. First, to an excellent approximation, the only possible shower constituents are electrons, positrons and photons and their interactions with matter reduce to pair creation in the case of photons and to bremsstrahlung in the case of electrons and positrons. Second, the shower development depends on a single scale, the atmospheric depth, measured in radiation lengths. 

At UHECR energies, most of the shower energy is therefore contained in electromagnetic electron-photon showers that have split away from the hadronic development process at the successive generations of interactions of the secondaries with the atmosphere. At variance with neutral pions, charged pions will either decay, in which case they will generate a muon component, or continue to interact with the atmosphere and therefore contribute to the further development of the hadronic shower.
 
The form used to parameterise the longitudinal profile of electromagnetic showers is the standard Gaisser-Hillas function \cite{9}:
\begin{equation}
S=S_{max}{(\frac{X-X^*}{X_{max}-X^*})}^{\frac{X_{max}-X^*}{w}} e^{\frac{X_{max}-X}{w}}  		      
\end{equation}
where $S$ is the density of charged particles at depth $X$ in the medium. In practice, $SdX$ may be the sum of the charged particle track lengths in the transverse shower slice between $X$ and $X+dX$, or the energy ionisation loss in that same slice, or even the amount of Cherenkov light produced in that same slice. At high energies, all three distributions are expected to have very similar shapes. The depth variable $X$ is measured in g/cm$^2$ with $dX$ being the product of the local density by the thickness of the slice. In atmospheric air the dependence of density on altitude distorts $X$ with respect to actual distances.
 
The quantity $X^*$ defines where the shower, understood as its charged particle components, starts developing. In the case of a photon, it starts at the location of the first pair creation while in the case of an electron it starts at $X^*=0$. Obviously, once started, the shower develops independently from $X^*$ and $S$ depends explicitly on $X-X^*$.
 
It is therefore sufficient to consider showers induced by electrons, {\it i.e.} having $X^*=0$. For such showers, the knowledge of $\langle X \rangle$ and of $Rms(X)$ fixes $w$ and $X_{max}$, that of $\Sigma=\int SdX$  fixes $S_{max}$. Explicitly \cite{8},

$$\delta=(\frac{\langle X \rangle}{Rms(X)})^2-1 \hspace{1cm} X_{max}=\frac{\langle X \rangle\delta}{\delta+1}$$
\begin{equation}
S_{max}=\frac{\Sigma \delta^{\delta+1}e^{-\delta}}{\Gamma(\delta+1)X_{max}} \hspace{1cm} w = \frac{X_{max}}{\delta}
\end{equation}

The dependence on energy of the mean and rms values of $\langle X \rangle$ and of $\rho=Rms(X)/\langle X \rangle$ have been parameterised once for all \cite{7} and are used in the present model to describe the longitudinal profile of the showers induced by the decay photons of neutral pions. The geometry of the decay is described exactly and the values of $X^*$ are chosen at random with an $exp(-\frac{7}{9}X^*/X_0)$ distribution.
 
Because of shower to shower fluctuations, the parameters that describe the average profile (obtained as superposition of a large number of different showers) are not the same as the mean values of the parameters that describe individual profiles. More precisely, the mean value of the former profile, $\langle X' \rangle$, and that of the mean values of the latter profiles, $\langle \langle X \rangle \rangle$, are equal and can be parameterized as $3.22+2.34lgE$. But the $\rho$ parameter of the former profile, $\rho'$, and the mean value of the $\rho$ parameters of the latter profile, $\langle \rho \rangle$, differ. In the case of the latter profiles, the rms values of the quantities $\langle X \rangle$ and $\rho$ define the size of the shower to shower fluctuations. To a very good approximation, $Rms(\langle X \rangle)$ is constant and equal to $0.94 \pm 0.01$ radiation lengths. On the contrary, $Rms(\rho)$ is found to decrease with energy as $Rms(\rho)\sim 0.001+16.20(lgE+5.6)^{-3}$. 

Photon showers have a lateral extension characterized by the Moli\`{e}re radius, $R_M$. To a good approximation, $R_M$ is an energy-independent constant equal to the radiation length multiplied by 21 MeV and divided by the critical energy \cite{10}. Therefore it scales with the radiation length, namely with the reciprocal of the atmospheric pressure. As the atmospheric pressure depends on altitude, it varies during shower development. However, in practice, we can retain the value on ground to be a good approximation in the description of the lateral distribution function, namely of the energy density on ground. A form $1/(R_{eff}^2+r^2)^2$ gives a good description of the global lateral extension of the energy density on ground, $r$ being the distance to the shower axis. The radius $R_{eff}$ has been adjusted in such a way that the energy deposited outside a cylinder of radius equal to the Moli\`{e}re radius be $\sim 10\%$ of the primordial photon energy \cite{10}. 
The result is $R_{eff} \sim R_M/3 \sim 20$ m.

\subsection{Hadronic interactions: an introduction}
The main feature of hadronic interactions is the peculiar distribution of the produced secondaries in phase space: a uniform distribution in rapidity and a steeply falling distribution in transverse momentum. L. Van Hove was first to state it explicitly \cite{11} and to introduce the concept of what he called ``longitudinal phase space'', the transverse momentum limitation having a scale given by the Planck constant $\hbar$ divided by the proton radius, $\sim$1 fm, namely of the order of 200 MeV/c. In the limit of infinite momentum, the invariance of a uniform rapidity distribution under Lorentz transformations implies that there exists no privileged momentum frame. Feynman was first to suggest a relation between such behaviour and a field theory of elementary hadron constituents that he called partons \cite{12} and which were later identified with gluons. Indeed, QCD \cite{13} reduces the strong interaction to essentially three Lagrangian terms associated with the bremsstrahlung-like radiation of gluons, either from a quark or from a gluon (in the form of triple and quadruple couplings), the latter being the result of the non-abelian nature of the theory. However, while these terms are easily accessible to experiments that probe short distances, implying the production of large transverse momenta, their effects are hidden at large distances: low transverse momentum interactions, such as those that prevail in the development of extensive air showers, can only rely on so-called ``QCD inspired'' approximate models \cite{14}.

Most of what is known today of the properties of hadronic interactions was learned in the late seventies and early eighties, in particular with experiment UA5 \cite{15} that is a reference in the field. In addition to the longitudinal phase space configuration, it includes:

$-$ the slow increase with energy of the total cross-section \cite{16};

$-$ the existence of diffractive events, where one of the protons is excited, its debris being separated in rapidity from central production;

$-$ the existence of short range rapidity correlations, well described in terms of clusters, of which only part are resonances \cite{17};

$-$ the existence of a leading effect, implying that the largest rapidity particle essentially carries the quantum numbers of the initial proton. Subtracting the leading energy and introducing accordingly the concept of effective energy \cite{18} for central production gives evidence for the universality of hadronization processes taking place in different interactions, such as electron-positron, lepton-nucleon and proton-proton collisions.

A phenomenological synthesis of experimental knowledge guided by QCD-inspired concepts is at the basis of all existing Monte Carlo simulations of extensive air shower development, including the model presented here. The universal features listed above, together with the requirement of energy-momentum conservation, leave fortunately little freedom to the modelist and it is not surprising that all models that respect such constraints produce similar results.

\subsection{Hadronic interactions: the model}
The general picture is that which emerges from the considerations developed in the preceding sub-section: two leading particles, each taking some 25\% of the available centre-of-mass energy, separated from a central rapidity plateau by two rapidity gaps. The rapidity plateau is characterized by a rather uniform density distribution and important short range rapidity correlations that are well described by clusters. These are seen in charge as well as in rapidity and transverse momentum. Transverse momentum distributions are steeply falling, first exponentially as expected from the Fourier transform of a disk, and later as a power law as expected from interacting point like constituents.
  
At variance with standard codes, the inelasticity is taken as an adjustable parameter and the parameters used to describe central production are calculated from the effective energy rather than from the total centre-of-mass energy.
 
The general algorithm used in the code is as follows \cite{7}:

a) Choose the fractions $\eta_1$ and $\eta_2$ of the centre-of-mass half-energy, $\sqrt{s}/2$, carried by the leading particles at random with Gaussian distributions having a mean value of 0.6 and an rms value of 0.15. The forward leading particle retains the identity of the projectile and the backward leading particle is simply ignored. The centre-of-mass energy available for central production, or effective energy \cite{18}, is therefore $\sqrt{s_{eff}}=(1-\eta_1-\eta_2)\sqrt{s}$. The leading particles do not carry any transverse momentum, and so do therefore globally the central secondaries, the longitudinal centre-of-mass momentum and energy of which are now defined.

b) Depending on $\sqrt{s_{eff}}$, choose the number of central clusters and the numbers of pions in each cluster in such a way as to reproduce the desired multiplicity distribution. Once this is done choose the width of the rapidity plateau in such a way as to conserve energy. Clusters are then distributed evenly at equal intervals on the plateau. A final adjustment of the cluster momenta is made to fine tune energy momentum conservation.
  
A library of clusters containing between two and seven pions is created. The transverse momentum distribution of the pions is chosen to reproduce that desired for central pions, the clusters being given no transverse momentum of their own. While the width of the rapidity plateau and the cluster rapidity density increase linearly with $lgs_{eff}$, implying that the cluster multiplicity increases  quadratically with $lgs_{eff}$, the number of pions per cluster and the transverse momentum distribution are nearly constant, increasing only slightly with $lgs_{eff}$.
 
The forms given to the pion transverse momentum distribution and to the mean values of the total and charged multiplicity distributions are given in Reference 7. The number of pions per cluster is chosen at random between 2 and 7 with a Gaussian distribution having a mean value of $1.6+0.21 lns_{eff}$ and an rms value of 1. The total number of clusters $n_{cl}$ is chosen at random with an {\it ad hoc} distribution meant to properly reproduce the final multiplicity distribution. Pions are defined to be charged or neutral at random according to experimental observations.
 
Figure 1 illustrates some of the above features and compares distributions of the present model with those of the HDPM model \cite{5}.
 
There exists no exact treatment of nuclei interactions. A standard approach, which is used here, is that of the Glauber model \cite{19}. A first useful concept is that of wound nucleons: when two nuclei collide, only some of their nucleons interact. These are defined as having their projection on a plane normal to the incident momentum contained within the intersection of the projections of both nuclei on the same plane. The calculation is straightforward once the nucleon radius and the Woods-Saxon distribution of nucleons inside the nuclei are known. The interaction of wounded nucleons is treated as a cascade of each of the projectile nucleon on the set of wound nucleons that are on its path. Details of the calculation are given in Reference 7.
 
Finally, pion-nucleon interactions are treated the same way as nucleon- nucleon interactions apart from the values taken by the interaction cross section which are taken from the HDPM model \cite{5} and updated using recent results from LHC \cite{6} and from the PAO \cite{20}. Precisely, the following parameterisations of the inelastic interaction cross-sections as functions of incident energy, $E_{inc}$, are used:
 
Nucleon nucleon: $lg\sigma [mb]= 1.340+0.067 lgE_{inc} [GeV]$

Nucleon air: $lg\sigma [mb]= 2.332+0.040 lgE_{inc} [GeV]$

Iron air: $lg\sigma [mb]= 3.197+0.022 lgE_{inc} [GeV]$

\begin{figure}
\centering
\begin{tabular}{cc}
\multicolumn{2}{c}{\includegraphics[width = 0.35\textwidth]{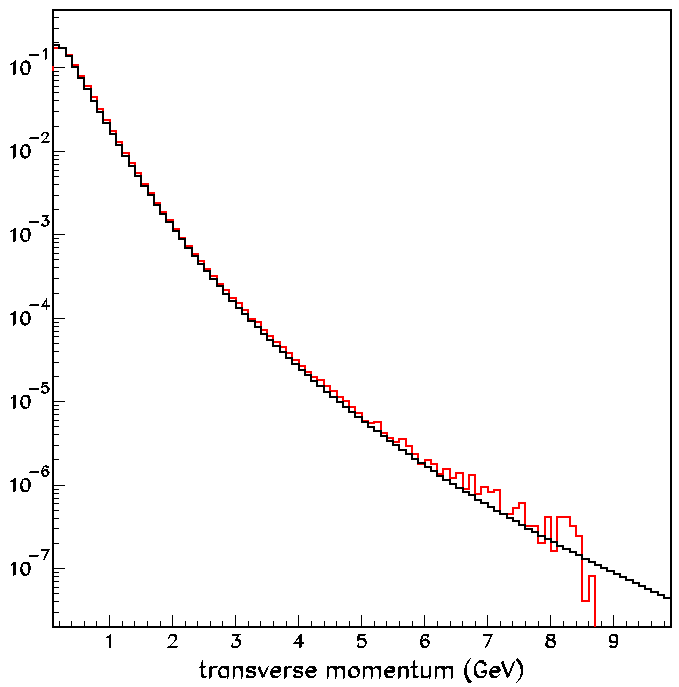}}\\
\includegraphics[width = .35\linewidth]{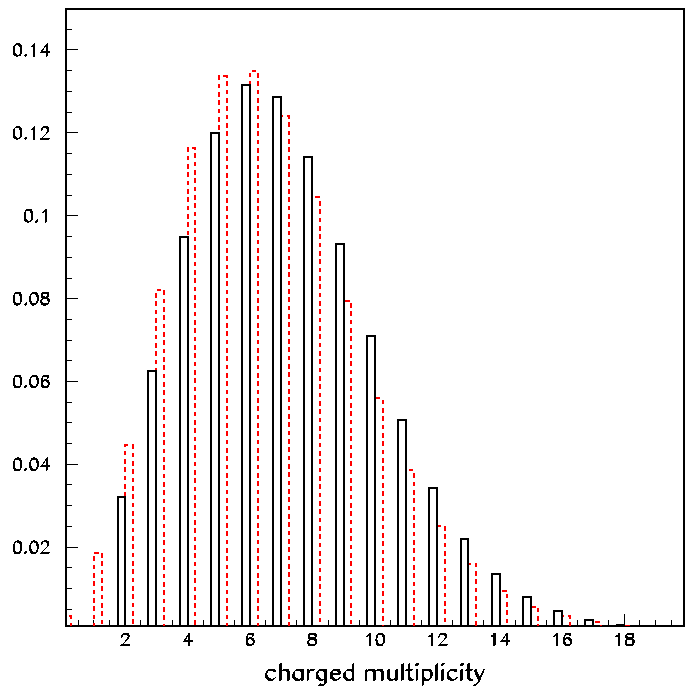} &
\includegraphics[width = .35\linewidth]{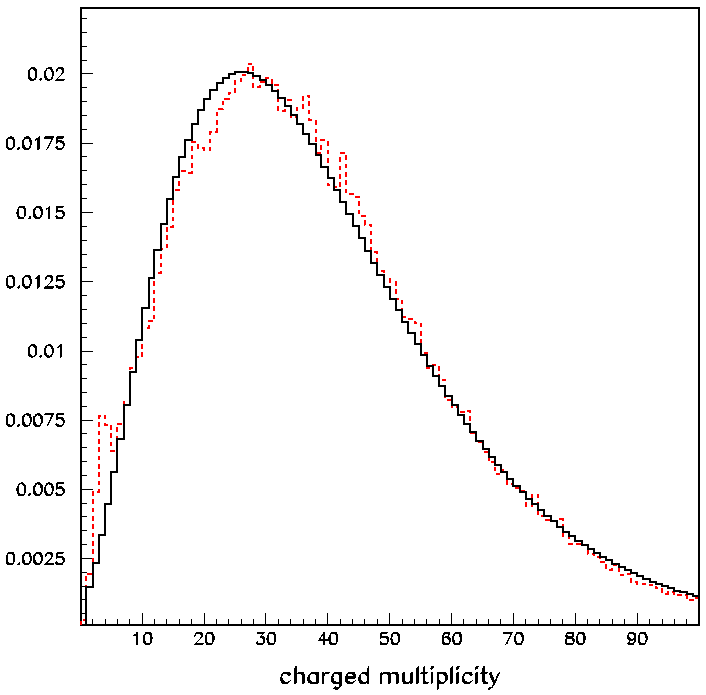} \\
\includegraphics[width = .35\linewidth]{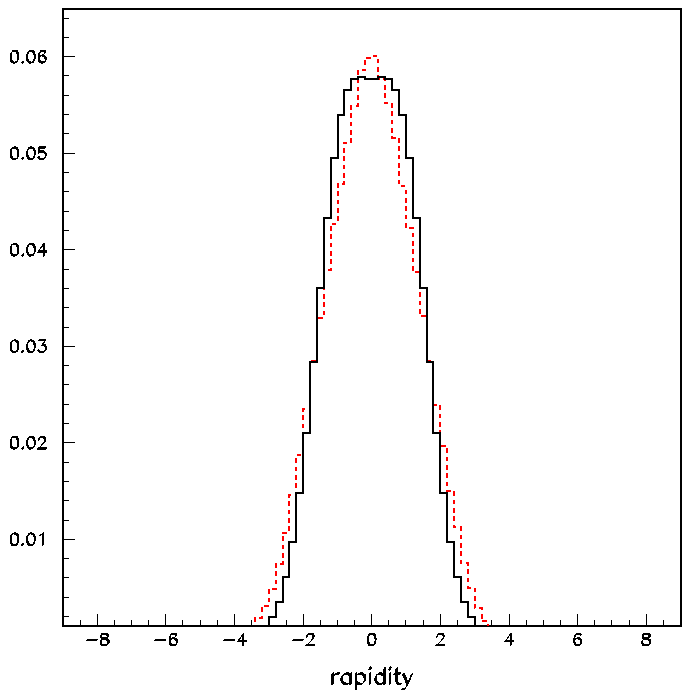} &
\includegraphics[width = .36\linewidth]{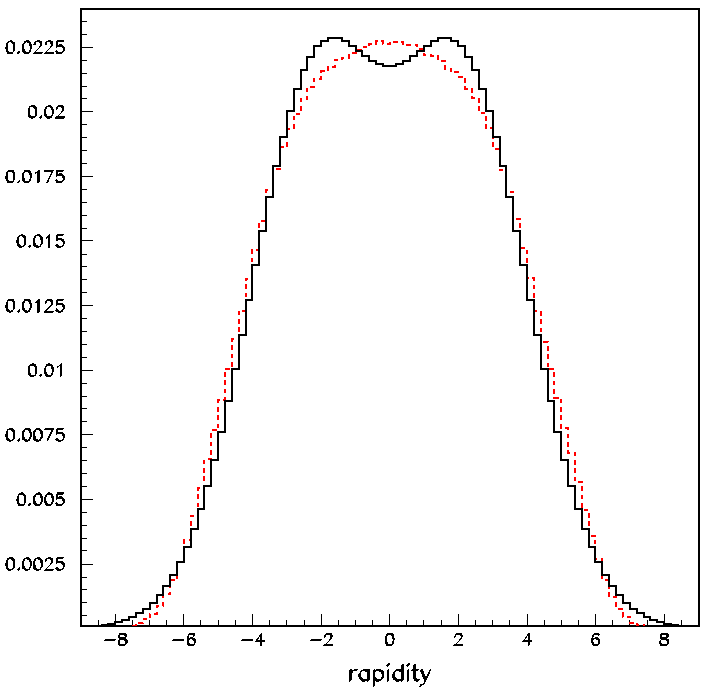}
\end{tabular}
\caption{Upper panel: Transverse momentum distribution: the result of the code (histogram) is compared with the analytical form (line) $dN/dp_t\approx (p_t/p_0)/(1+p_t/p_0)^{10}$ with $p_0=1.47$ GeV. Middle panels: Charged multiplicity distributions obtained here at 10$^2$ GeV (left) and 10$^6$ GeV (right) are compared with those of the HDPM model \cite{5}. Lower panels: Pion rapidity distributions obtained here at 10$^2$ GeV (left) and 10$^6$ GeV (right) are compared with those of the HDPM model \cite{5}.}
\end{figure}

\subsection{Parameterisations}
The aim is to obtain parameterisations of three profiles as a function of three variables: the profiles are the longitudinal shower profile, the muon lateral distribution function and the electron/photon lateral distribution function; the variables are associated with the primary: they are its energy, the altitude of its first interaction and the cosine of its zenith angle of incidence. It is sufficient to limit the parameterisation to pion induced sub-showers: at each step of the shower development, the interaction of the leading nucleon with atmosphere is treated separately using the hadronic interaction model.

Neutral pions are made to decay into two photons that are immediately converted into parameterised sub-showers. Charged pions are made to decay or to interact according to the relative values taken by the decay length or interaction length. If they interact, the treatment they are given depends on the value of the ratio between their energy and the primordial energy: if it is smaller than a fraction $f$ of the primary energy, they are converted into a parameterised sub-shower and if it is larger, the hadronic interaction model is used to describe the interaction. If they decay, they are simply converted into a muon according to the proper kinematics. Electrons from muon decays are ignored: the muons are simply removed from the set of shower particles once they have decayed. As the transverse momentum distribution of decay muons in the pion rest frame is invariant, the lateral scale of the muon lateral distribution function is proportional to altitude above ground and inversely proportional to momentum. However, multiple scattering, energy loss and occasional muon decays break this simple scaling law and smear the transverse distribution. Detailed descriptions for vertical and oblique incidences are given in Reference 7.
 
 The longitudinal profile is measured along the shower axis defined as the primary momentum and may extend to very large depths, well beyond ground, the assumption being that atmospheric pressure keeps increasing according to the same exponential law as in the real atmosphere. The reason is to guarantee a sensible Gaisser-Hillas parameterisation of the profile, which requires performing the fit well beyond shower maximum. However, in the case of the transverse profile, the energy contained in the shower when it reaches ground is fully distributed in the lateral distribution function. The charged pion and muon contributions to the longitudinal profile are ignored: we only retain that of electromagnetic showers resulting from neutral pion decays, however normalized to their energies. The lateral distribution functions are given in the plane normal to the shower axis at its intersection with ground. Obtaining the measured signal requires a projection on ground and a simulation of the detector response. The muon lateral distribution function is given in muons per square meters and the electron/photon lateral distribution function is given in MeV per square meters.

\begin{figure}
\centering
\begin{tabular}{cc}
\includegraphics[width = .38\textwidth]{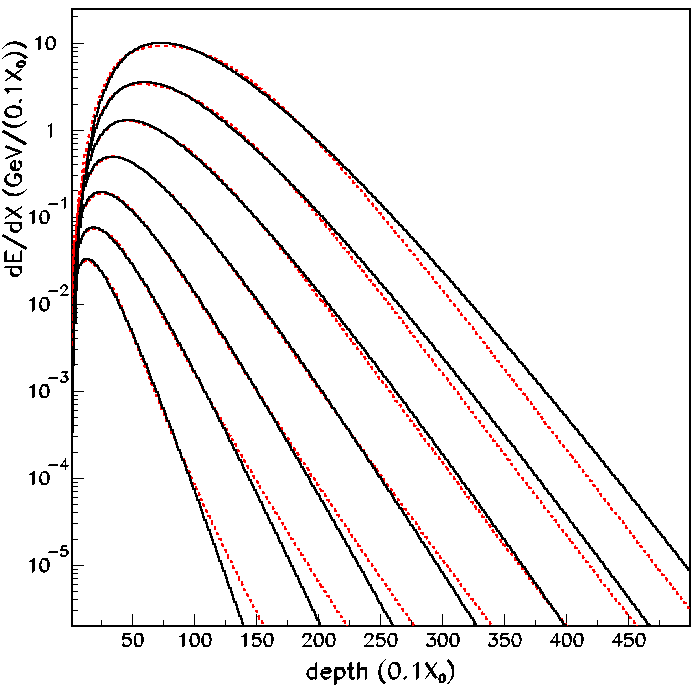} &
\includegraphics[width = .38\linewidth]{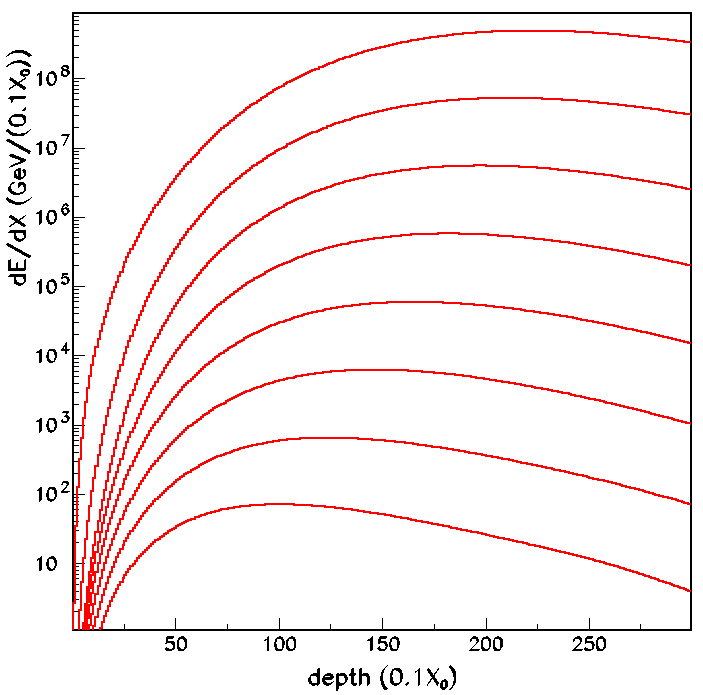}\\
\includegraphics[width = .38\linewidth]{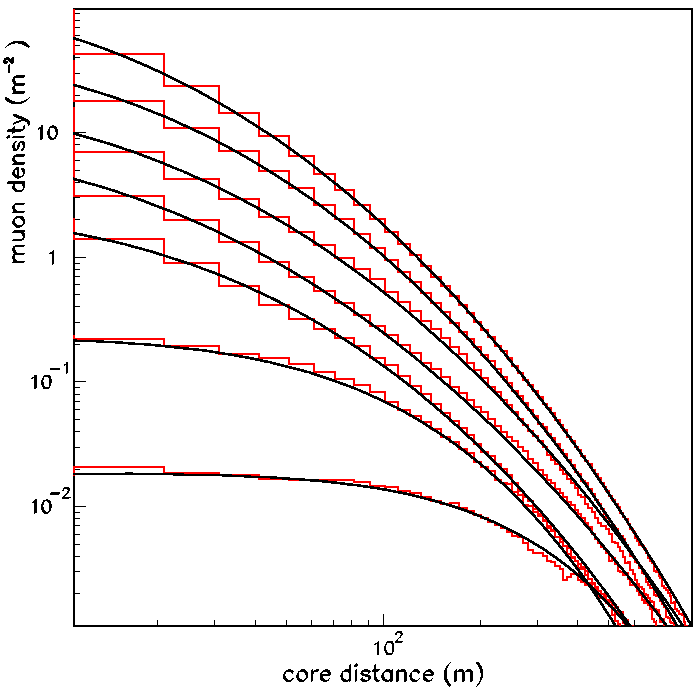} &
\includegraphics[width = .38\linewidth]{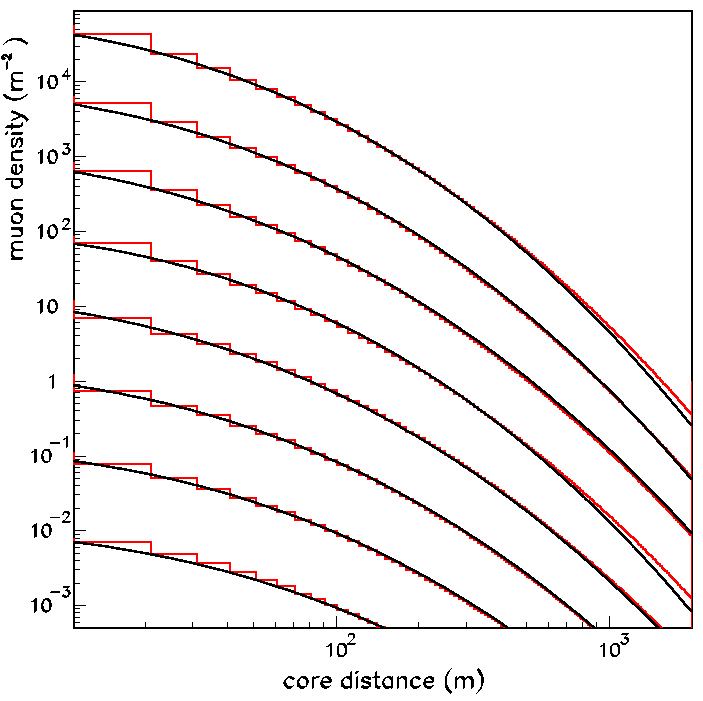} \\
\includegraphics[width = .38\linewidth]{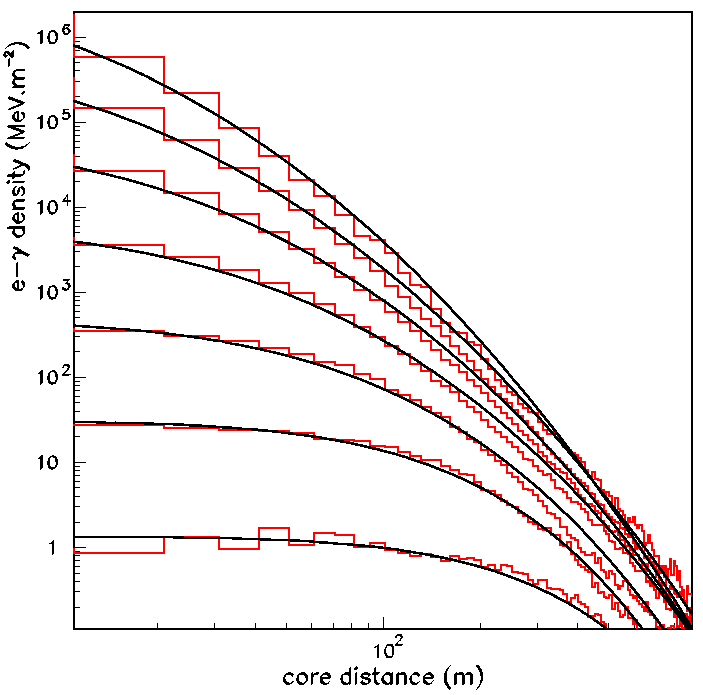} &
\includegraphics[width = .38\linewidth]{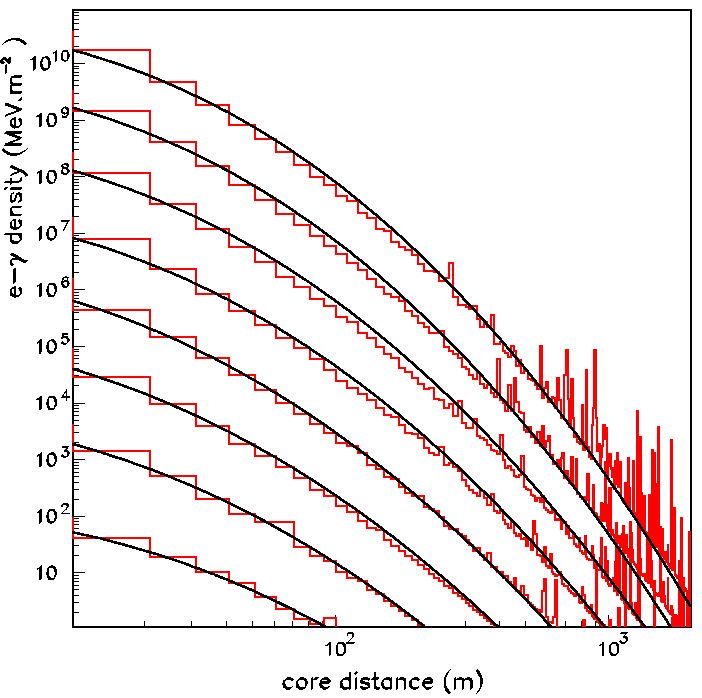}
\end{tabular}
\caption{Parameterized shower profiles induced by a vertical charged pion: longitudinal profiles (upper), muon lateral distribution function (middle) and electron-photon lateral distribution functions (lower). In the left panels the pion decays 1 km above ground at seven energies (1, 3.2, 10, 32, 100, 316 and 1000 GeV). In the right panels, the pion decays at 22 km above ground at energies in geometrical progression between 10 TeV and 0.1 ZeV.}
\end{figure}

In a first phase, a grid is chosen in the parameter space that scans from 1 GeV to 1 TeV, from $cos\theta=0.5$ to $cos\theta=1$ and from $z=0$ to $\sim$22 km above ground. At each node of the grid lattice, 10000 showers are generated and the three profiles are parameterised. The parameters of a new shower are then calculated by interpolation. In practice, linear interpolations are used for $cos\theta$ and logarithmic interpolations for energy and altitude. 

In a second phase, parameterisation is extended to higher energies stepwise, by successive iterations. In order to keep manageable computing time, showers induced by pions having energy smaller than $f=5\%$ of the primary energy are no longer simulated but simply replaced by parameterised showers. This allows extending the parameter grid up to $10^{20}$ eV. As a parameterised average shower replaces each sub-shower, shower-to-shower fluctuations are the exclusive result of fluctuations in the sample of interactions having incident energies exceeding the fraction $f$ of the primary energy, here taken equal to 5\%.
   
Updating the longitudinal profile is done by directly adding the new sub-shower profile to the already accumulated main shower profile, starting from the depth at which the interaction occurs. However, updating the lateral distribution functions cannot be done so simply: the sub-shower lateral distribution function is parameterized as a function of distance $r_{sub}$ to the sub-shower axis but its contribution to the main shower must be in terms of the distance $r_{main}$ to the main shower axis, and $r_{sub}$ and $r_{main}$ are not related by a simple analytic form. What is done in practice is to choose 100 values of $r_{sub}$ at random, each with a weight of 1\%, and for each of these add the proper contribution to the $r_{main}$ distribution. The form used to parameterize the lateral distribution functions is $exp[a+b(lnr)^c]$ with the distance $r$ measured in meters and the lateral distribution function evaluated in MeV/m$^2$ for electron-photons and in muons/m$^2$ for muons.

Typical parameterisations are illustrated in Figure 2 where the direct results of the simulation are compared with the parameterisations that are made of them.
  
\section{Results and discussion}
\subsection{Elongation rates}
Figure 3 illustrates the results obtained for $\langle X_{max} \rangle$ and $Rms(X_{max})$ for proton and iron primaries at incident energies between 10$^{18}$ and 10$^{20}$ eV. They are compared with experimental data \cite{4} and with some of the predictions obtained from standard simulation codes \cite{5}. The general agreement is quite remarkable given the crudeness of the present model. The results (in g/cm$^{2}$) can be parameterized as follows as a function of the primary energy $E$ (in eV):

$\langle X_{max} \rangle _p=806(\pm2)+41(\pm3)[lgE-19]$

$\langle X_{max} \rangle _{Fe}=716.5(\pm0.7)+56.3(\pm0.8)[lgE-19]$

$Rms(X_{max})_p=61.6(\pm1.3)-5.7(\pm2.0)[lgE-19]$

$Rms(X_{max})_{Fe}=19.0(\pm0.2)-0.8(\pm0.3)[lgE-19]$

The uncertainties given in parentheses are statistical only.

\begin{figure}[h]
  \begin{minipage}[b]{0.48\linewidth}
    \centering
    \includegraphics[width = 1.\linewidth]{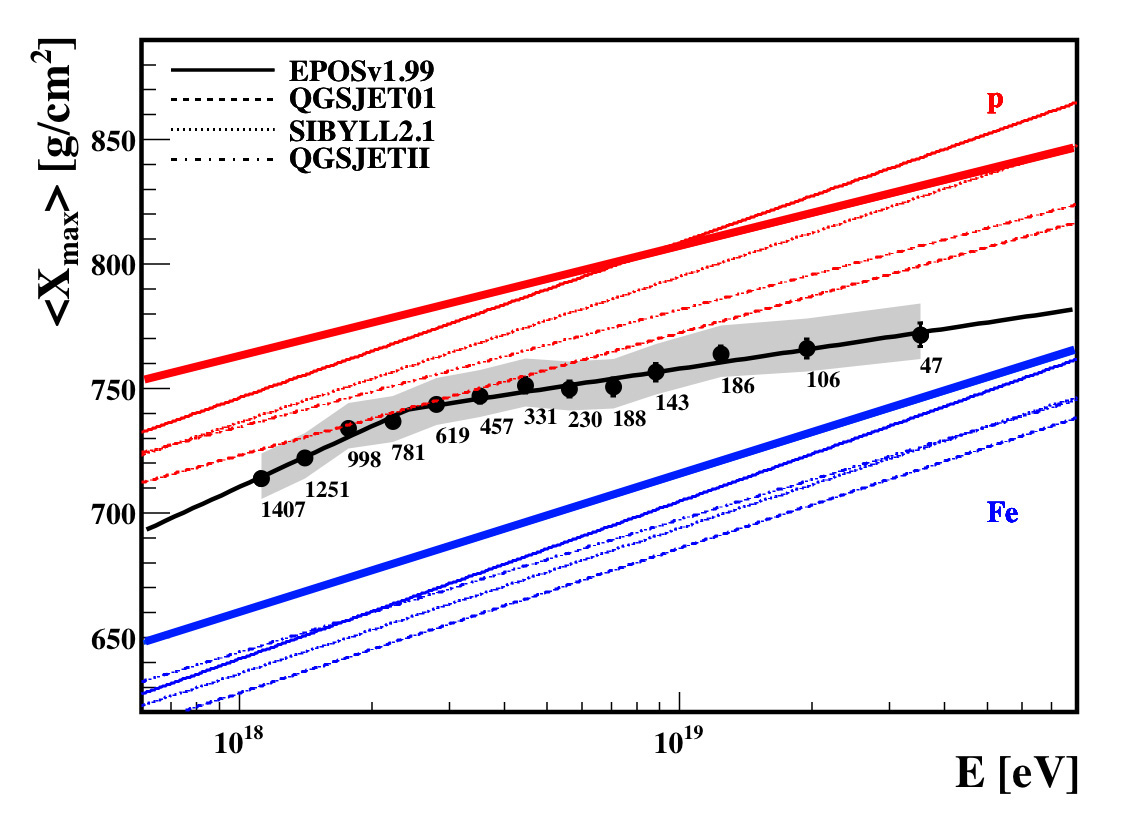}
  \end{minipage}
  \hspace{0.1cm}
  \begin{minipage}[b]{0.48\linewidth}
    \centering
    \includegraphics[width = 1.\linewidth]{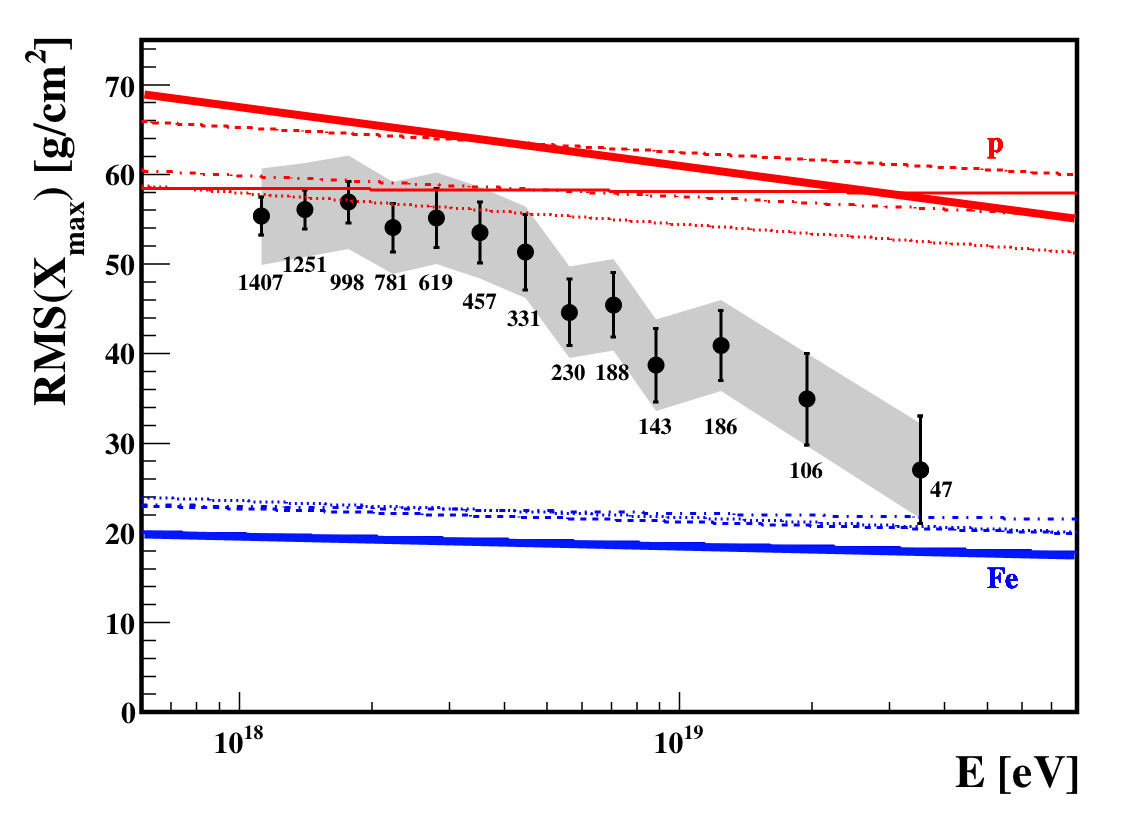}
  \end{minipage}
  \caption{Recently measured $\langle X_{max} \rangle$ and $Rms(X_{max})$ distributions \cite{4} are compared with the predictions of the present model (thick lines) and of conventional models \cite{5} (thin lines).}
\end{figure} 

The above results are given for a zenith angle incidence of 60$^o$ in order to make sure that the shower is well beyond maximum when reaching ground. In the simulation, incidence is largely irrelevant in that respect but in real data care is normally taken to consider only showers having a well visible maximum.

The smooth dependence on energy of the above quantities is commonly measured in terms of absolute elongation rates defined for $\textless X_{max} \textgreater$ and $Rms(X_{max})$ as the increase of the relevant variable by decade of energy: $ER_{mean}=\partial \langle X_{max} \rangle /\partial lgE$ and $ER_{rms}=\partial Rms(X_{max})/ \partial lgE$. Similarly, relative elongation rates may be defined as $ER_{mean}=\partial \langle X_{max} \rangle /(\langle X_{max} \rangle \partial lgE)$ and $ER_{rms}=\partial Rms(X_{max})/[Rms(X_{max})\partial lgE]$. Recently measured \cite{21} elongation rates at 10$^{19}$ eV are given in Reference 22. Their relative values are $3.6\pm1.5\%$ for HiRes, $4.5\pm2.5\%$ for Yakutsk, $3.4\pm0.7\%$ for Auger and $4.1\pm2.3\%$ for the Telescope Array, namely an average of $3.6\pm0.5\%$ corresponding to an absolute elongation rate of $27\pm4$ g/cm$^{2}$ while conventional models \cite{5,22} predict typically twice as much for protons, corresponding to a relative elongation rate of 6\% to 8\%.  The relative elongation rates predicted by the present model (Table 1, line $a$) are $5.1\pm0.4\%$ for protons and $7.9\pm0.1\%$ for iron, in reasonable agreement with, but slightly lower than the predictions of conventional models.

The low value of the measured elongation rates, when compared with model predictions, may be interpreted in terms of an increase of the average primary mass when energy approaches the GZK suppression. However, as thoroughly discussed in Reference 22, it would be premature to reach such a conclusion when the experimental situation is not yet fully settled. There exist strong arguments in favour of a proton-iron dominance in this energy range, intermediate mass nuclei being photo-dissociated on their way to Earth \cite{23}. Attempts at interpreting the data in terms of a simple proton-iron mixture have triggered considerations on a possible unexpected energy dependence of the strong interaction in this energy domain \cite{24}. However, blaming the observed change of the elongation rate on some new phenomenon would imply rather drastic revisions of our understanding of the standard model interactions as no new threshold is expected in this energy range. Moreover, such a threshold in the parton-parton centre-of-mass system would be significantly smeared in the Earth rest frame. Yet, it is interesting to study the dependence of the above quantities on some parameters of the model in order to better quantify the statements that have just been made.

The dependences on elasticity and on the neutral-to-charged ratio have been evaluated at a fixed primary energy of 10$^{19}$ eV with the following results:

As a function of elasticity $q_{mean}$, one obtains in the range 0.5$<q_{mean}<$0.7:

$\langle X_{max} \rangle _p=808(\pm1)+80(\pm16)[q_{mean}-0.6]$

$\langle X_{max} \rangle _{Fe}=719.1(\pm0.5)+38(\pm6)[q_{mean}-0.6]$

$Rms(X_{max})_p=62.9(\pm1.5)+49(\pm21)[q_{mean}-0.6]$

$Rms(X_{max})_{Fe}=19.0(\pm0.2)+3.1(\pm3.1)[q_{mean}-0.6]$.

These are very small effects: changing $q_{mean}$ by 20\% of its value changes $\langle X_{max} \rangle$ by only 1\% or less of its own value.

In the UHECR energy range, the neutral-to-charged ratio, $N/C$, used in the model is essentially constant and equal to 0.545. Changing it between 0.45 and 0.65 one obtains:

$\langle X_{max} \rangle _p=807(\pm2)-35(\pm22)[N/C-0.545]$

$\langle X_{max} \rangle _{Fe}=717.5(\pm0.1)-15.4(\pm1.1)[N/C-0.545]$

$Rms(X_{max})_p=61.7(\pm1.9)-25(\pm27)[N/C-0.545]$

$Rms(X_{max})_{Fe}=19.1(\pm0.4)-9.5(\pm5.1)[N/C-0.545]$.

Here again, the effect is small: changing $N/C$ by 20\% of its value changes $\langle X_{max} \rangle$ by less than a percent of its own value.

The predicted elongation rates are listed in Table 1 for three different input parameters: 

a) inelasticity $q_{mean}=0.6$ and neutral-to-charged ratio $N/C=0.545$,

b) inelasticity $q_{mean}=0.7$ and neutral-to-charged ratio $N/C=0.545$,

c) inelasticity $q_{mean}=0.6$ and neutral-to-charged ratio $N/C=0.65$.

They are in qualitative agreement with predictions made under similar conditions using conventional models \cite{22,24}.

\begin{table}[h]
  \caption{Relative elongation rates predicted by the present model (see text).}
  \centering
  \begin{tabular}{|c|c|c|c|c|}
    \hline
    \multirow{2}{*}{} & \multicolumn{2}{|c|}{Proton primary} & \multicolumn{2}{|c|}{Iron primary} \\ \cline{2-5}
    & $ER_{mean}$ & $ER_{rms}$& $ER_{mean}$ & $ER_{rms}$ \\ 
    \hline
    a)&$5.1\pm0.4\%$&$-9.3\pm3.2\%$&$7.9\pm0.1\%$&$-4.2\pm1.6\%$\\
    \hline
    b)&$4.9\pm0.1\%$&$-9.5\pm2.5\%$&$7.6\pm0.1\%$&$-9.0\pm1.0\%$\\
    \hline
    c)&$5.6\pm0.1\%$&$-1.9\pm1.0\%$&$7.8\pm0.1\%$&$-2.8\pm2.8\%$\\
    \hline 
    \end{tabular}
\end{table}

The above predictions have been obtained with an $f$ value of 0.1\%, implying that inelasticity and neutral-to-charged ratio calculated at an incident energy of 10$^{19}$ eV are not modified below 10$^{16}$ eV. The authors of References 22 and 24 assume instead a progressive change of the parameters with energy, more realistic than the abrupt change considered here. However, the point being made here is simply to illustrate qualitatively the robustness of the predicted elongation rates when the model parameters are varied within reasonable limits: for such a task, the present approach is sufficient.

\subsection{Lateral distribution functions}
Lateral distribution functions (LDF) have been simulated together with the longitudinal shower profiles. As illustrated in Figure 4, iron and proton primaries are observed to display very similar dependences on the distance to the shower axis. As mentioned earlier, LDFs are evaluated in the plane normal to the shower axis at its intersection with ground. A comparison with observed LDFs requires a projection of this plane on ground and a simulation of the detector response, which is beyond the scope of the present study. Both are expected to smear the LDF and to make it less steep. 

\begin{figure}[ht]
  \centering
  \includegraphics[width = 0.4\textwidth]{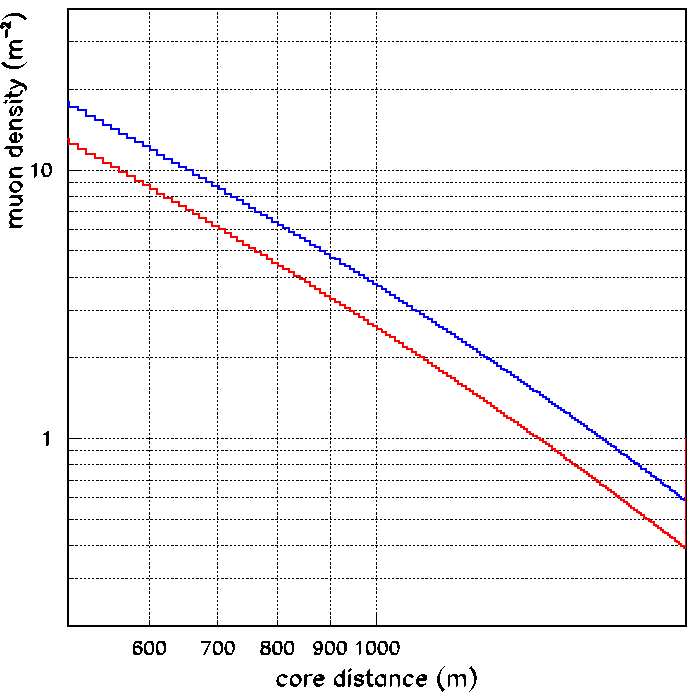}
  \caption{Muon LDF of iron (above) and proton (below) showers at 10$^{19}$ eV.}
  \label{fig:figure4}
\end{figure}

The ratio of the muon yields on ground between iron and proton primaries is expected to be largely independent from the precise rate of decrease of the LDF with distance. It is predicted to be $R_{\mu}=1.31\pm0.01$ in standard conditions (case $a$) of Table 1; at vertical incidence, this number becomes $R_{\mu}=1.49\pm0.01$. This result is in good agreement with the predictions of standard simulation codes [5]. The dependences on energy, elasticity and neutral-to-charged ratio are given below for 60$^o$ incidence: 

$R_{\mu}=1.32(\pm0.01)+0.024(\pm0.018) [lgE-19]$

$R_{\mu}=1.33(\pm0.01)-0.31(\pm0.11)[q_{mean}-0.6]$

$R_{\mu}=1.33(\pm0.01)+0.76(\pm0.11) [N/C-0.545]$.

For each of the proton and iron LDFs, the total muon yields, $Y_{\mu}$, increase slightly less than in proportion with energy. Precisely, writing $Y_{\mu}=E\Xi_{\mu}$, $\partial \Xi_{\mu}(p)/[\Xi_{\mu}(p)\partial lgE]=-16.5\pm1.5\%$ and $\partial \Xi_{\mu}(Fe)/[\Xi_{\mu}(Fe)\partial lgE]=-14.8\pm0.1\%$ at 60$^o$ incidence. At 0$^o$ incidence, these numbers are $-23.2\pm0.7\%$ and $-23.1\pm0.5\%$ respectively. Both proton- and iron-induced muon yields are predicted to have very similar energy dependences, resulting in a nearly energy-independent value of $R_{\mu}$. 

Both the present model and currently available simulation codes \cite{25} predict iron-induced muon yields $-$and {\it a fortiori} proton-induced muon yields$-$ significantly lower than experimentally observed. The average pion transverse momentum used in the present model is smaller than that of kaons and more massive mesons in actual interactions. The predicted muon yield would become consistent with that measured if the distance scale were expanded by some 25\%. As it is essentially governed by the value of the average transverse momentum, increasing the latter by the same amount at all energies would achieve the desired result. However, such an increase is only efficient at intermediate energies, corresponding to intermediate altitudes that dominate the muon density on ground and increasing the average transverse momentum at higher energies exclusively would not help. This illustrates the difficulty to predict muon yields in agreement with observation.

The electron/photon LDF shows qualitatively similar features as the muon LDF. In particular proton-induced and iron-induced energy densities on ground have similar energy dependences resulting in a nearly energy-independent ratio, $R_{e\gamma}=0.48(\pm0.01)+0.00(\pm0.01)[lgE-19]$ at 60$^o$ and $R_{e\gamma}=0.682(\pm0.004)+0.084(\pm0.006)[lgE-19]$ at vertical incidence.

\subsection{Concluding remarks}
A simple method of simulation of the development of ultra high energy extensive air showers has been presented. Its satisfactory performance has been illustrated on several examples. The method, which is based on an algorithm of parameterisation of lower energy sub-showers, does not have the ambition to compete with existing sophisticated models \cite{5,24} but provides a useful complement to their predictions. The results presented here have illustrated the robustness of the elongation rates and of the iron to proton ratios predicted by sensible models of the interaction of cosmic rays with the Earth atmosphere in the ultra high energy domain. They illustrate how little freedom there is in the description of the interactions of ultra high energy cosmic rays with the Earth atmosphere: these must obey the severe constraints of energy-momentum conservation and of longitudinal phase-space, leaving little freedom to the modelist in the absence of a new threshold. The same comment applies to the lateral distribution functions, the main features of which are properly reproduced. The ability of such a crude model to make sensible predictions is quite remarkable but does not come as a real surprise when one considers the very general arguments that govern the physics of shower development \cite{22} and when one remembers the predictive powers of much cruder models, such as proposed by Heitler and Matthews \cite{26}.
 
Interpreting the observed data in terms of new phenomena in the interaction of cosmic rays with the Earth atmosphere seems premature in the present experimental situation: in the absence of a new threshold, simple scaling properties are expected to be obeyed with the most relevant scales being the radiation and interaction lengths and the pion decay length. On the contrary, the proximity of the GZK suppression offers a natural scale that can be expected to cause significant changes in the mass composition. 

\section*{Acknowledgements}
We express our deepest gratitude to our colleagues in the Pierre Auger Collaboration for their constant interest and invaluable support. One of us (P.N.D.) thanks Dr F. Fleuret for clarifications on using the Glauber model. Financial and/or material support from the Institute for Nuclear Studies and Technology, National Foundation for Science and Technology Development (NAFOSTED), the World Laboratory, Rencontres du Vietnam and Odon Vallet fellowships is gratefully acknowledged. 
%% References

\bibliographystyle{elsarticle-num}
\bibliography{<your-bib-database>}

\end{document}